\documentclass[12pt,a4]{article}
\usepackage{amsmath}
\usepackage{bm}
\usepackage{amsfonts}
\usepackage{amssymb}
\usepackage{graphics}
\usepackage[normal]{caption2}
\usepackage{subfigure}
\usepackage{rotating}
\usepackage{citesort}
\def\be{\begin{equation}}
\def\ee{\end{equation}}
\def\ba{\begin{array}}
\def\ea{\end{array}}
\def\bea{\begin{eqnarray}}
\def\eea{\end{eqnarray}}

\begin{document}
\baselineskip 20pt \setlength\tabcolsep{2.5mm}
\renewcommand\arraystretch{1.5}
\setlength{\abovecaptionskip}{0.1cm}
\setlength{\belowcaptionskip}{0.5cm}
\pagestyle{empty}
\newpage
\pagestyle{plain} \setcounter{page}{1} \setcounter{lofdepth}{2}
\begin{center} {\large\bf N/Z and N/A dependence of balance energy as a probe of symmetry energy in heavy-ion collisions}\\
\vspace*{0.4cm}
{\bf Aman D. Sood}\footnote{Email:~amandsood@gmail.com}\\
{\it  SUBATECH, Laboratoire de Physique Subatomique et des
Technologies Associ\'{e}es, Universit\'{e} de Nantes - IN2P3/CNRS
- EMN 4 rue Alfred Kastler, F-44072 Nantes, France.\\}
\end{center}

We study the N/Z and N/A dependence of balance energy (E$_{bal}$) for
isotopic series of Ca having N/Z (N/A) varying from 1.0 to 2.0 (0.5 to 0.67). We show
that the N/Z (N/A) dependence of E$_{bal}$ is sensitive to symmetry
energy and its density dependence at densities higher than
saturation density and is insensitive towards the isospin
dependence of nucleon-nucleon (nn) cross section and Coulomb repulsion.
We also study the effect of momentum dependent interactions (MDI) on the N/Z (N/A) dependence of E$_{bal}$. We find
that although MDI influences the E$_{bal}$ drastically, the N/Z (N/A) dependence of E$_{bal}$ remains unchanged on inclusion of MDI.


\newpage
\baselineskip 20pt
\section{Introduction}
With the availability of high-intensity radioactive beams at many
facilities as well as a number of next generation beam facilities
being constructed or being planned \cite{rib1,rib2}, the studies
on the role of isospin degree of freedom have recently attracted a
lot of attention in both nuclear physics and astrophysics. The
ultimate goal of such studies is to extract information on the
isospin dependence of in-medium nuclear effective interactions as
well as equation of state (EOS) of isospin asymmetric nuclear
matter. The later quantity especially the symmetry energy term is
important not only to nuclear physics community as it sheds light
on the structure of radioactive nuclei, reaction dynamics induced
by rare isotopes but also to astrophysics community as it acts as
a probe for understanding the evolution of massive stars and
supernova explosion \cite{marr10}. It is worth mentioning that the
equation of state of symmetric nuclear matter has been constrained
up to densities 5 times the normal nuclear matter density through
the measurements of transverse flow as well as its disappearance
along with other collective flows (like radial flow, elliptic
flow) \cite{daniel02} and of subthreshold kaon production in
relativistic nucleus-nucleus collisions \cite{fuch06}.
\par
 Although the nuclear symmetry energy at normal nuclear matter
density is known to be around 30 MeV \cite{li02}, its values at
other densities are poorly known. Heavy-ion collisions induced by
radioactive beams provide unique opportunities to investigate the
isospin-dependent properties of asymmetric nuclear matter,
particularly the density dependence of symmetry energy
\cite{li98}. Experimentally symmetry energy is not a directly
measurable quantity and has to be extracted from observables
related to symmetry energy. Over the last decade a large number of
experimental observables have been proposed like neutron/proton
ratio of emitted nucleons \cite{li97}, the neutron-proton
differnetial flow \cite{li00}, the t/$^{3}$He \cite{chen303},
$\pi^{-}/\pi^{+}$ \cite{li02,gait04}, $\Sigma^{-}/\Sigma^{+}$
\cite{qli205}, and K$^{0}/K^{+}$ \cite{ferini06} ratios and so on.
A recent analysis of data has led to a symmetry energy term of the
form
 E$_{sym}$ $\simeq$
31.6($\frac{\rho}{\rho_{0}})^{\gamma}$ MeV with $\gamma$ =
0.4-1.05 for densities between 0.1$\rho_{0}$ and 1.2$\rho_{0}$
\cite{shetty10}. However, for all the above mentioned observables
the Coulomb force of charged particles plays an important role. It
competes strongly with symmetry energy. Recently Gautam and Sood
\cite{gaum10} studied the relative contribution of Coulomb force
and symmetry energy in isospin effects on the collective
transverse flow as well as its disappearance for isobaric systems
throughout the mass range and colliding geometry. They clearly
demonstrated the dominance of Coulomb repulsion over the symmetry
energy. The collective transverse in-plane flow disappears at a
particular energy called as balance energy \cite{gaum10,krof89}.
In recent communication, Gautam \emph{et al}. \cite{gaum210} has
studied the transverse momentum for a neutron rich system
$^{60}Ca$+$^{60}Ca$ in the Fermi energy as well as at high
energies. There they find that transverse momentum is sensitive to
the symmetry energy as well as its density dependence in the Fermi
energy region. Motivated by those results we here study the
E$_{bal}$ as a function of N/Z and N/A of the system for an isotopic
series. We here choose the isotopes so that the Coulomb repulsion
is same for all the systems, since as mentioned previously that
Coulomb plays much dominant role as compared to symmetry energy in
isospin effects. Here we will demonstrate that the N/Z (N/A) dependence
of E$_{bal}$ for the isotopes of same element is a sensitive probe
for the symmetry energy as well as its density dependence. To
check the sensitivity of N/Z (N/A) dependence of E$_{bal}$ towards
density dependence of symmetry energy, we have calculated the
E$_{bal}$ throughout the isotopic series for different forms of
symmetry energy F$_{1}(u)$, F$_{2}(u)$, and F$_{3}(u)$,
where\emph{ u} = $\frac{\rho}{\rho_{0}}$. The different forms are
described later. The present study is carried out within the
framework of Isospin-dependent Quantum Molecular Dynamics
(IQMD)\cite{hart98} Model. Section 2 describes the model in brief.
Section 3 explains the results and discussion and section 4
summarizes the results.
\section{The model}
The IQMD model treats different charge states of nucleons, deltas
and pions explicitly, as inherited from the
Vlasov-Uehling-Uhlenbeck (VUU) model. The IQMD model has been used
successfully for the analysis of a large number of observables
from low to relativistic energies. One of its versions QMD model
has been quite successful in explaining various phenomena such as
multifragmentation \cite{kumar}, collective flow \cite{sood1}, and
hot and dense nuclear matter \cite{leh1} as well as particle
production \cite{huang}. The isospin degree of freedom enters into
the calculations via symmetry potential, cross sections and
Coulomb interaction.
 \par
 In this model, baryons are represented by Gaussian-shaped density distributions
\begin{equation}
f_{i}(\vec{r},\vec{p},t) =
\frac{1}{\pi^{2}\hbar^{2}}\exp(-[\vec{r}-\vec{r_{i}}(t)]^{2}\frac{1}{2L})
\times \exp(-[\vec{p}- \vec{p_{i}}(t)]^{2}\frac{2L}{\hbar^{2}})
 \end{equation}
 Nucleons are initialized in a sphere with radius R = 1.12 A$^{1/3}$ fm, in accordance with liquid-drop model.
 Each nucleon occupies a volume of \emph{h$^{3}$}, so that phase space is uniformly filled.
 The initial momenta are randomly chosen between 0 and Fermi momentum ($\vec{p}$$_{F}$).
 The nucleons of the target and projectile interact by two- and three-body Skyrme forces, Yukawa potential, Coulomb interactions,
  and momentum-dependent interactions (MDI). In addition to the use of explicit charge states of all baryons and mesons, a symmetry potential between protons and neutrons
 corresponding to the Bethe-Weizsacker mass formula has been included. The hadrons propagate using Hamilton equations of motion:
\begin {eqnarray}
\frac{d\vec{{r_{i}}}}{dt} = \frac{d\langle H
\rangle}{d\vec{p_{i}}};& & \frac{d\vec{p_{i}}}{dt} = -
\frac{d\langle H \rangle}{d\vec{r_{i}}}
\end {eqnarray}
 with
\begin {eqnarray}
\langle H\rangle& =&\langle T\rangle+\langle V \rangle
\nonumber\\
& =& \sum_{i}\frac{p^{2}_{i}}{2m_{i}} + \sum_{i}\sum_{j>i}\int
f_{i}(\vec{r},\vec{p},t)V^{ij}(\vec{r}~',\vec{r})
 \nonumber\\
& & \times f_{j}(\vec{r}~',\vec{p}~',t) d\vec{r}~ d\vec{r}~'~
d\vec{p}~ d\vec{p}~'.
\end {eqnarray}
 The baryon potential\emph{ V$^{ij}$}, in the above relation, reads as
 \begin {eqnarray}
  \nonumber V^{ij}(\vec{r}~'-\vec{r})& =&V^{ij}_{Skyrme} + V^{ij}_{Yukawa} +
  V^{ij}_{Coul} + V^{ij}_{mdi} + V^{ij}_{sym}
    \nonumber\\
   & =& [t_{1}\delta(\vec{r}~'-\vec{r})+t_{2}\delta(\vec{r}~'-\vec{r})\rho^{\gamma-1}(\frac{\vec{r}~'+\vec{r}}{2})]
   \nonumber\\
   &  & +t_{3}\frac{\exp(|(\vec{r}~'-\vec{r})|/\mu)}{(|(\vec{r}~'-\vec{r})|/\mu)}+
    \frac{Z_{i}Z_{j}e^{2}}{|(\vec{r}~'-\vec{r})|}
   \nonumber \\
   &  & +t_{4}\ln^{2}[t_{5}(\vec{p}~'-\vec{p})^{2} +
    1]\delta(\vec{r}~'-\vec{r})
    \nonumber\\
   &  & +t_{6}\frac{1}{\varrho_{0}}T_{3i}T_{3j}\delta(\vec{r_{i}}~'-\vec{r_{j}}).
 \end {eqnarray}
Here \emph{Z$_{i}$} and \emph{Z$_{j}$} denote the charges of
\emph{ith} and \emph{jth} baryon, and \emph{T$_{3i}$} and
\emph{T$_{3j}$} are their respective \emph{T$_{3}$} components
(i.e., $1/2$ for protons and $-1/2$ for neutrons). The
parameters\emph{ $\mu$} and \emph{t$_{1}$,....,t$_{6}$} are
adjusted to the real part of the nucleonic optical potential.
 For the density dependence of  the nucleon optical potential, standard Skyrme-type parametrization is employed.
   We also use the isospin and energy-dependent cross
  section  $\sigma$ = 0.8 $\sigma_{nn}^{free}$.
The details about the elastic and inelastic cross sections for
proton-proton and proton-neutron collisions can be found in
\cite{hart98,cug}. The cross sections for neutron-neutron
collisions are assumed to be equal to the proton-proton cross
sections. Explicit Pauli blocking is also included; i.e. Pauli
blocking of the neutrons and protons is treated separately. We
assume that each nucleon occupies a sphere in coordinate and
momentum space. This trick yields the same Pauli blocking ratio as
an exact calculation of the overlap of the Gaussians will yield.
We calculate the fractions P$_{1}$ and P$_{2}$ of final phase
space for each of the two scattering partners that are already
occupied by other nucleons with the same isospin as that of
scattered ones. The collision is blocked with the probability
\begin {equation}
 P_{block} = 1-[1 - min(P_{1},1)][1 - min(P_{2},1)],
\end {equation}
and, correspondingly is allowed with the probability 1 -
P$_{block}$. For a nucleus in its ground state, we obtain an
averaged blocking probability $\langle P_{block}\rangle$ = 0.96.
Whenever an attempted collision is blocked, the scattering
partners maintain the original momenta prior to scattering.
 The different forms of symmetry energy are obtained by changing
the density dependence of the potential part of the symmetry
energy (last term in eq. (4)). The various forms are F$_{1}(u)
\propto u$, F$_{2}(u) \propto u^{0.4}$, F$_{3}(u) \propto u^{2}$
(where \emph{ u} = $\frac{\rho}{\rho_{0}}$). F$_{4}$ represents
calculations without symmetry potential.
  \par
   \section{Results and discussion}
 We have simulated several thousand events at incident energies around
 balance energy in small steps of 10 MeV/nucleon for each
isotopic system of Ca+Ca having N/Z (N/A) varying from 1.0 to 2.0 (0.5-0.67). i.e.
Ca$^{40}$+Ca$^{40}$, Ca$^{44}$+Ca$^{44}$,
Ca$^{48}$+Ca$^{48}$, Ca$^{52}$+Ca$^{52}$,
Ca$^{56}$+Ca$^{56}$, and Ca$^{60}$+Ca$^{60}$ for
the semicentral colliding geometry range of 0.2 - 0.4. Such
systematic studies performed at low incident energies using
various fusion models have shown a linear enhancement in the
fusion probabilities with neutron content \cite{puri1}. We use
soft equation of state with and without MDI, labeled respectively
as Soft and SMD. The calculations with this choice of equation of
state and cross section were in good agreement with the data
throughout the colliding geometry \cite{gaum310}. The IQMD model
has also been able to reproduce the other data (example, high
energy proton spectra, gamma production) at incident energies
relevant in this paper \cite{ger98,gamm}. The reactions are
followed till the transverse flow saturates. The saturation time
is around 100 fm/c for the systems in the present study. For the
transverse flow, we use the quantity "\textit{directed transverse
momentum $\langle p_{x}^{dir}\rangle$}" which is defined as
\cite{sood,leh}
\begin {equation}
\langle{p_{x}^{dir}}\rangle = \frac{1} {A}\sum_{i=1}^{A}{sign\{
{y(i)}\} p_{x}(i)},
\end {equation}
where $y(i)$ is the rapidity and $p_{x}$(i) is the momentum of
$i^{th}$ particle. The rapidity is defined as
\begin {equation}
Y(i)= \frac{1}{2}\ln\frac{{\textbf{{E}}}(i)+{{\textbf{p}}}_{z}(i)}
{{{\textbf{E}}}(i)-{{\textbf{p}}}_{z}(i)},
\end {equation}
where ${\textbf{E}}(i)$ and ${\textbf{p}_{z}}(i)$ are,
respectively, the energy and longitudinal momentum of $i^{th}$
particle. In this definition, all the rapidity bins are taken into
account. A straight line interpolation is used to calculate the
E$_{bal}$. It is worth mentioning that the E$_{bal}$ has the same
value for all fragments types \cite{west93,pak97,west98,cuss}.

\begin{figure}[!t] \centering
 \vskip 0.5cm
\includegraphics[angle=0,width=10cm]{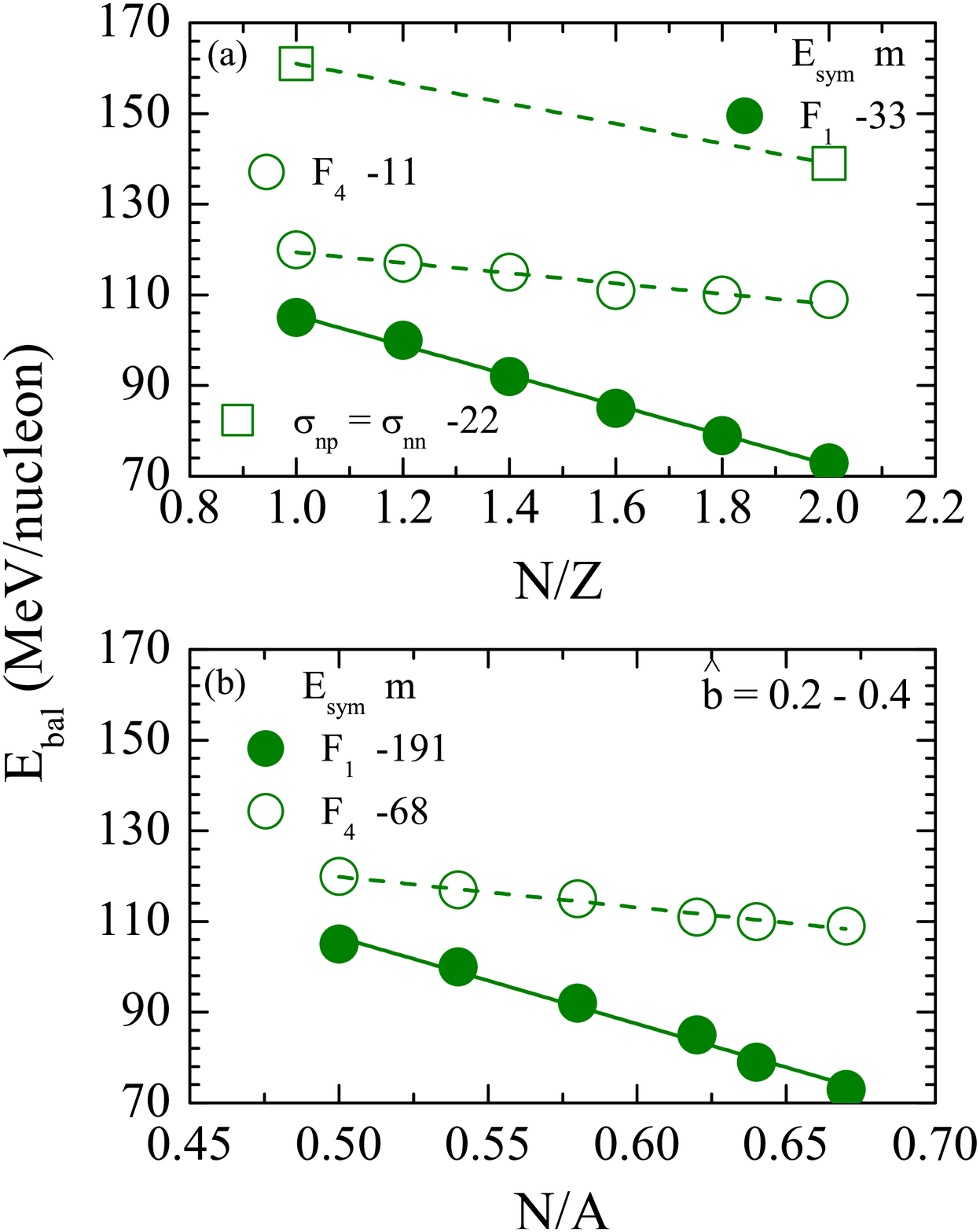}
 \vskip 0.5cm
\caption{(Color online) E$_{bal}$ as a function of N/Z (upper panel) and N/A (lower panel) of system
for E$_{sym} \propto F_{1} (u)$ and F$_{4}$. Lines are linear fit proportional to m. Various symbols are
explained in the text.}\label{fig3}
\end{figure}

 In fig. 1(a) we display the E$_{bal}$ as a function of
N/Z of the system. Solid green circles represent the calculated
E$_{bal}$. Lines are the linear fit to E$_{bal}$. We see that
E$_{bal}$ follows a linear behavior $\propto$ m$\ast$N/Z. As the
N/Z of the system increases, the mass of the system increases due
to addition of neutron content. In addition, the effect of
symmetry energy also increases with increase in N/Z. To check the
relative contribution of increase in mass with N/Z and symmetry
energy towards the N/Z dependence of E$_{bal}$, we make the
strength of symmetry energy zero and calculate E$_{bal}$. The
results are displayed by open circles in fig. 1(a). E$_{bal}$
 again follows a linear behavior $\propto$ m$\ast$N/Z. However, E$_{bal}$ decreases very slightly with increase in
 N/Z, whereas when we include the symmetry energy
 also in our calculations then the $\mid m \mid$ increases by 3
 times which
 shows that N/Z dependence of E$_{bal}$ is highly sensitive
 to the symmetry energy. The slight decrease in the E$_{bal}$ with
 N/Z (for calculations without symmetry energy) is due to increase
 in number of nucleon-nucleon collisions. To further explore this
 point, we switch off the symmetry energy and also make the cross
 section isospin independent (i.e. $\sigma_{np}$ = $\sigma_{nn}$
 and calculate E$_{bal}$ for two extreme N/Z. The results are displayed in
 fig. 1(a) by open squares. Again E$_{bal}$ follows a linear
 behavior. We see that the E$_{bal}$ for both $^{40}$Ca +
 $^{40}$Ca and $^{60}$Ca + $^{60}$Ca increases as expected.
 However, the increase in E$_{bal}$ for system with N/Z = 1 is
 more as compared to the system with N/Z = 2. This is  because
 with increase in N/Z the neutron number increases due to which
 neutron-neutron and neutron-proton collisions pairs increase.
 However, the increase in number of neutron-neutron collision
 pairs is much larger as compared to neutron-proton collision
 pairs. Therefore, the possibility of neutron-proton collision is
 much less in system with N/Z = 2. That is why the effect of
 isospin dependence of cross section decreases with increase in
 N/Z.
\par
In fig. 1(b), we display the E$_{bal}$ as a function of N/A of the system. Symbols have same meaning as in fig. 1(a). Again E$_{bal}$ follows
a linear behaviour with m = -191 and -68, respectively, for F$_{1}$ (u) and F$_{4}$. However, the percentage difference $\Delta E_{bal}$ \% (where $\Delta E_{bal}$ \% =
$\frac{E_{bal}^{F_{1}(u)}-E_{bal}^{F_{4}}}{E_{bal}^{F_{1}(u)}}$) is same (about 65\%) in both the figs. 1(a) and 1(b) which
shows that the effect of symmetry energy is same whether we discuss in terms of N/Z or N/A.
  \begin{figure}[!t] \centering
 \vskip 0.5cm
\includegraphics[angle=0,width=16cm]{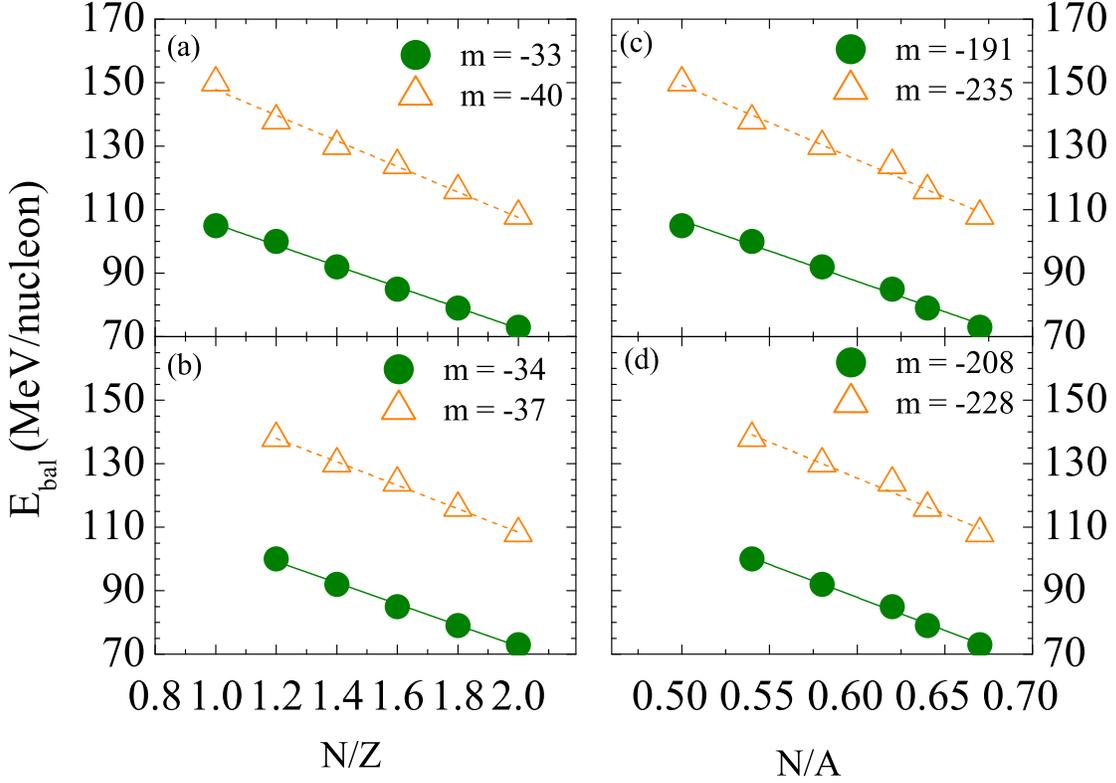}
 \vskip 0.5cm
\caption{(Color online) E$_{bal}$ as a function of N/Z (left panels) and N/A (right panels) of system
for isospin independent cross section for (a) and (c) N/Z = 1.0 to 2.0 (b) and (d)
N/Z = 1.2 to 2.0. Lines are linear fit proportional to m. Various symbols are explained in the
text.}\label{fig3}
\end{figure}

  As stated in literature, the isospin dependence
of collective flow and its disappearance has been explained as the
competition among various reaction mechanisms, such as, nn
collisions, symmetry energy, surface property of the colliding
nuclei and Coulomb force \cite{li}. Here we aim to show that the
N/Z and N/A dependence of E$_{bal}$ is sensitive to the symmetry energy
only. Since we are taking the isotopic
 series, the effect of Coulomb will be same for all the reactions.
  We have also checked that the N/Z dependence of E$_{bal}$ is insensitive to the
 EOS of symmetric nuclear matter. Moreover, as mentioned previously, the
 equation of state of symmetric nuclear matter has been constrained up to
 densities five times the normal matter density. In the present case as the N/Z of the system increases, the number of neutrons also increases.
 Since we are using isospin-dependent nn cross section, so to check the sensitivity of N/Z and N/A
 dependence of E$_{bal}$ to the isospin dependence of cross section, we calculate the E$_{bal}$ throughout
 the isotopic series by making the cross section isospin independent (fig. 2(a) open orange triangles, left panels).
 Again E$_{bal}$ follows a linear behavior with m = -40. We find
 that although E$_{bal}$ for individual system
 is very sensitive to the isospin dependence of cross section. However,
 N/Z dependence of E$_{bal}$ (for isotopic series) is much less sensitive to the isospin dependence of cross
 section.

  \begin{figure}[!t] \centering
 \vskip 0.5cm
\includegraphics[angle=0,width=10cm]{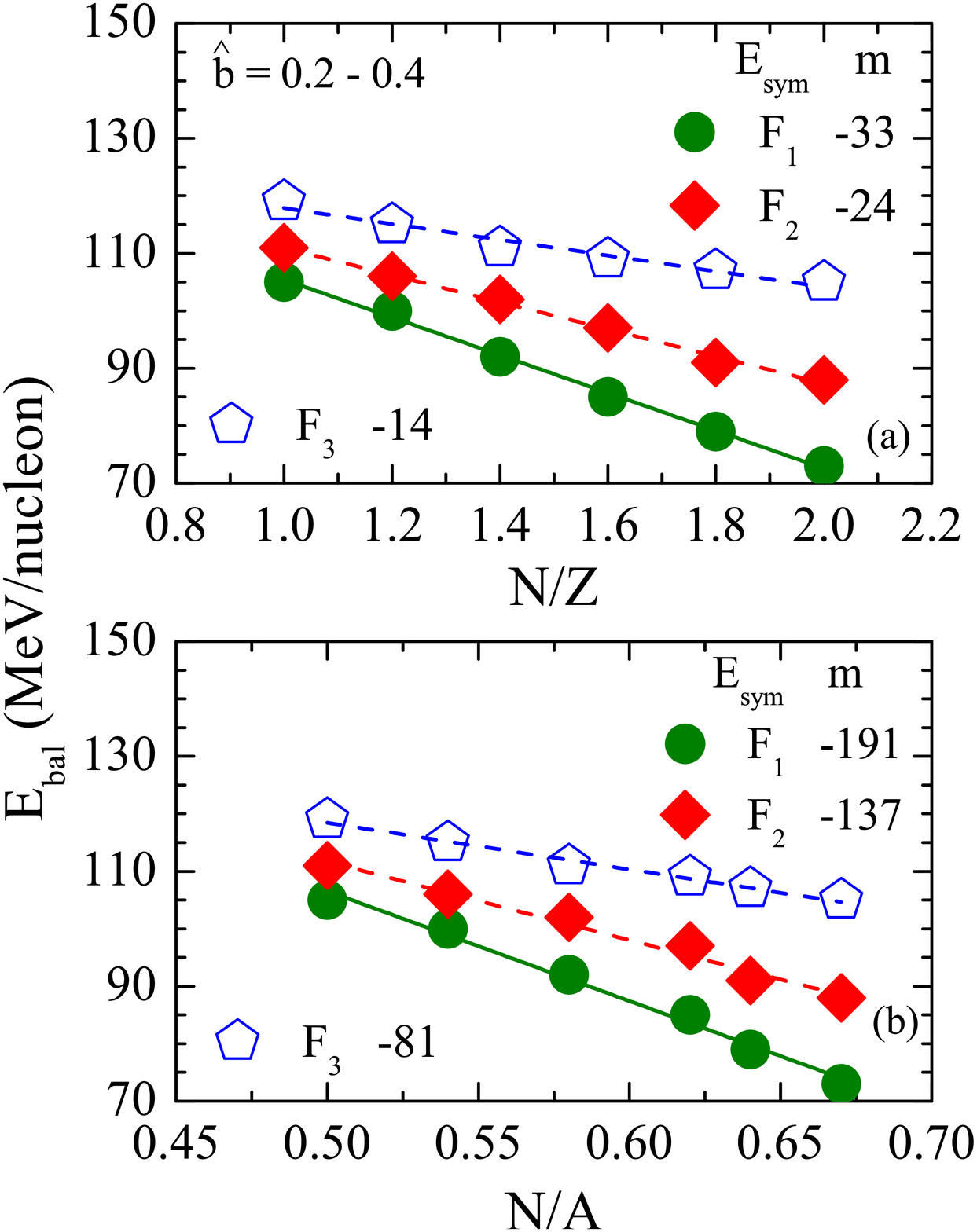}
 \vskip 0.5cm
\caption{(Color online) E$_{bal}$ as a function of N/Z (upper panel) and N/A (lower panel) of system
for E$_{sym} \propto F_{2} (u)$ and F$_{3} (u)$. Lines are linear fit proportional to m. Various symbols
are explained in the text.}\label{fig3}
\end{figure}

 \par
 In fig. 2(b) (left panels), we show E$_{bal}$ as a function of N/Z of the
 system for the N/Z range from 1.2 to 2.0. We find that the sensitivity of N/Z dependence of E$_{bal}$
 towards the isospin dependence of
cross section decreases further. Now m is -34 (-37) for
calculations with (without) isospin dependence of cross section.
Thus the N/Z dependence of E$_{bal}$ for neutron-rich isotopes is
sensitive only to symmetry energy. In figs. 2(c) and 2(d) (right panels) we display similar plots as in
the corresponding  left
panels but now we plot E$_{bal}$ as a function of N/A of the system. Again the percentage difference between the two curves in both
upper panels is same (about 18 \%) and same in lower panels as well (about 8 \%).
\par
 In fig. 3 (a) (3b) we display the N/Z (N/A) dependence of
E$_{bal}$ for different forms of symmetry energy; F$_{1}$(u)
(solid circles), F$_{2}$(u) (diamonds), and F$_{3}$(u)
(pentagons). For all the cases E$_{bal}$ follows a linear
behavior. Clearly, N/Z (N/A) dependence of E$_{bal}$ is sensitive to the
density dependence of symmetry energy as well. For a fixed N/Z (N/A) stiff symmetry energy F$_{1}$(u) shows
less E$_{bal}$ as compared to soft F$_{2}$(u) whereas super stiff
symmetry energy F$_{3}$(u) shows more E$_{bal}$ as compared to
F$_{2}$(u).
\begin{figure}[!t] \centering
 \vskip -1cm
\includegraphics[width=10cm]{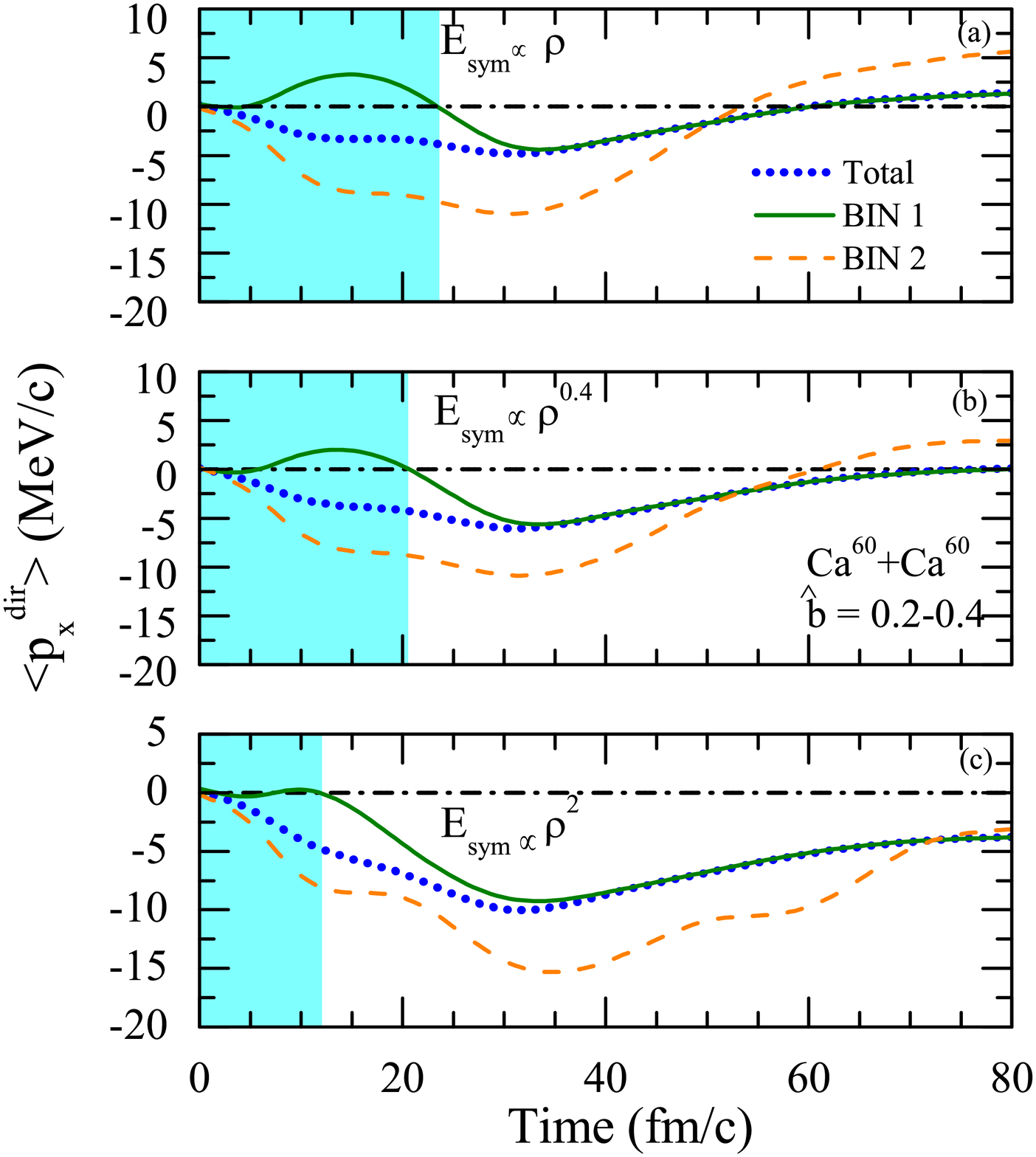}
\caption{(Color online) The time evolution of $<p_{x}^{dir}>$ for
different forms of symmetry energy for different bins at
b/b$_{max}$=0.2-0.4 . Lines are explained in the
text.}\label{fig2}
\end{figure}

To explain the above mentioned feature, we calculate for $^{60}$Ca
+ $^{60}$Ca the transverse flow of particles having $\rho/\rho_{0}
\leq 1$ (denoted as BIN 1) and particles with $\rho/\rho_{0} > 1$
(denoted as BIN 2), separately at all time steps for symmetry
energy F$_{1}$(u), F$_{2}$(u), and F$_{3}$(u). The incident energy
is taken to be 100 MeV/nucleon. The results are displayed in fig.
4. Solid (dashed) lines represent the p$_{x}^{dir}$ of particles
lying in BIN 1 (BIN 2). Dotted line represent the total
$<p_{x}^{dir}>$. We see that the total $<p_{x}^{dir} >$ is maximum
for stiff symmetry energy and minimum for super stiff symmetry
energy. During the initial stages of the reaction, $<p_{x}^{dir}
>$ due to particles lying in BIN 1 remains positive for F$_{1}$(u)
and F$_{2}$(u) because in the spectator region, repulsive symmetry
energy will accelerate the particles away from the overlap zone.
The effect is more pronounced for F$_{1}$(u) as compared to
F$_{2}$(u). Moreover, for F$_{1}$(u) and F$_{2}$(u), this interval
is about 5-25 fm/c and 5-20 fm/c, respectively.  This is because
although for F$_{2}$(u), the effective strength of symmetry energy
will be more for low density particles as compared to F$_{1}$(u),
however, in the central dense zone the effective strength of
F$_{2}$(u) will be less i.e. in the central dense zone, F$_{2}$(u)
will be less repulsive, therefore for F$_{2}$(u), there will be
more attractive force on the particles lying in the spectator
region towards the central dense zone as compared to that in case
of F$_{1}$(u). That is why during the initial stages the peak
value of $<p_{x}^{dir}
>$ as well as the duration for which it remains positive is less
for F$_{2}$(u) as compared to F$_{1}$(u) (compare shaded area in
fig. 4(a) and 4(b).
 This decides the value of $<p_{x}^{dir} >$ at saturation, which is more for F$_{1}$(u) as compared to F$_{2}$(u). In case of F$_{3}$(u) (fig. 4(c)) for particles lying in BIN 1,
i.e. ($\rho/\rho_{0} \leq 1$ ), the strength of symmetry energy
will be much smaller which is not sufficient to push the particles
away from the overlap zone. Therefore, the $<p_{x}^{dir} >$ of BIN
1 particles remains zero during the initial stages. This leads to
least value of final state $<p_{x}^{dir} >$ for super stiff
symmetry energy as compared to stiff and soft symmetry energy. The
$<p_{x}^{dir}>$ due to particles in BIN 2 (dashed line) decreases
in a very similar
 manner for all the different symmetry energies between 0-10 fm/c. Between 10-25 fm/c, $<p_{x}^{dir}>$ for
 F$_{3} (u)$ decreases more sharply as compared to in case of F$_{1} (u)$ and
F$_{2} (u)$. This is because in this time interval the particles
from BIN 1 enters into BIN 2 and $<p_{x}^{dir}>$ of particles
entering BIN 2 from BIN 1 in case of F$_{1} (u)$ and F$_{2} (u)$
will be less negative due to stronger repulsive symmetry energy as
compared to in case of F$_{3}$ (u) (see Ref. \cite{gaum210} also).
During the expansion phase, i.e. after 30 fm/c the total
$<p_{x}^{dir} >$ and $<p_{x}^{dir} >$ of BIN 1 particles overlap
as expected. Therefore, the effect of symmetry energy on the low
density particles during the initial stages decide the fate of the
final $<p_{x}^{dir} >$ and hence E$_{bal}$.

\begin{figure}[!t] \centering
 \vskip 0.5cm
\includegraphics[angle=0,width=16cm]{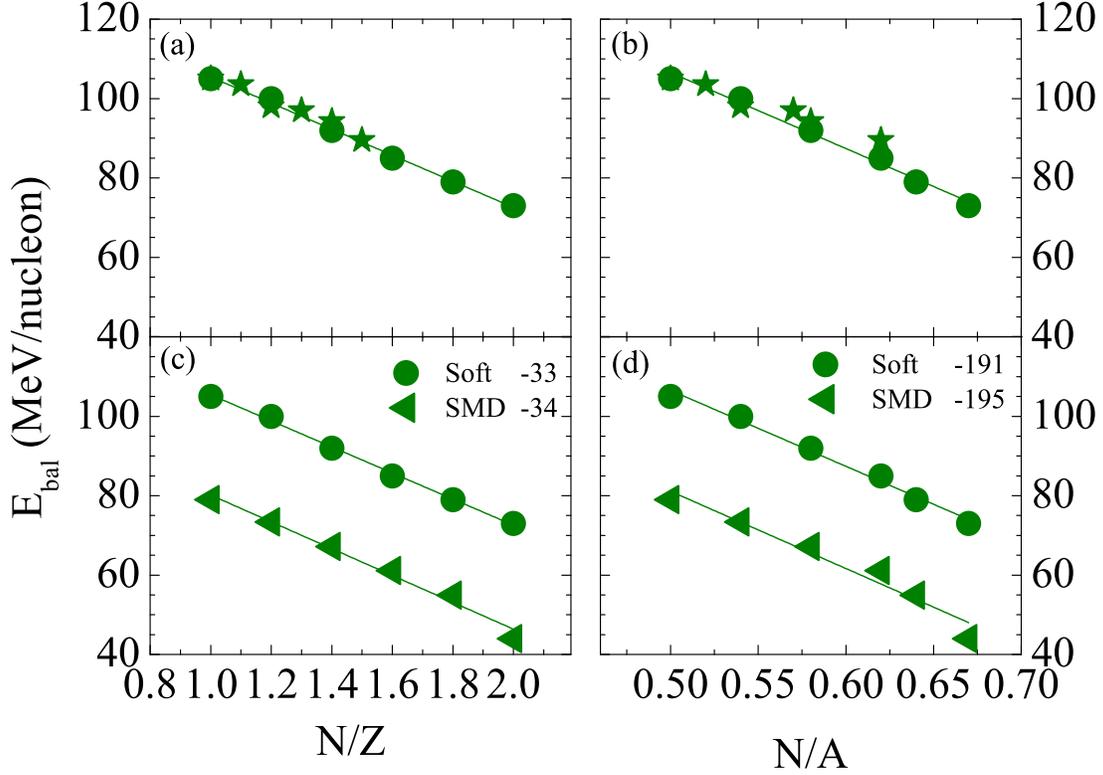}
 \vskip 0.5cm
\caption{ (a) and (b) (upper panels) E$_{bal}$ as a function of N/Z (left panel) and N/A (right panel) of system for E$_{sym}
\propto F_{1} (u)$ with $^{40}$Ca as
target (stars). (c) and (d) (lower panels) E$_{bal}$ as a function of N/Z (left panel) and N/A (right panel) of system for E$_{sym}
\propto F_{1} (u)$ with SMD EOS (left triangles). Circles represent the values of E$_{bal}$ as in fig. 1. Lines are linear fit proportional to m.}\label{fig3}
\end{figure}

\par
Since one cannot use radioactive isotopes as targets,
therefore, as a next step we fix the target as a stable isotope
$^{40}$Ca and vary the projectile from $^{40}$Ca  to $^{60}$Ca and
calculate E$^{bal}$. In this case the N/Z (N/A) of the reaction varies
between 1 to 1.5 (0.5 to 0.6) and the asymmetry $\delta = \frac {A_{1}-A_{2}}{A_{1}+A_{2}}$
of the reaction varies from 0
to 0.2. The results are displayed by solid green stars in figs. 5(a) and (b) (upper panels).
The solid green circles represent the calculations for symmetric
reactions with N/Z (N/A) varying from 1 to 2 (0.5 to 0.67), i.e. $^{40}$Ca+$^{40}$Ca
to $^{60}$Ca+$^{60}$Ca. Lines represent the linear fit $\propto$
m. We see that N/Z (N/A) dependence of E$_{bal}$ is same for both the
cases. We also find that when we use
stable target $^{40}$Ca  and radioactive target $^{60}$Ca,
the N/Z (N/A) decreases from 2 (0.67) in case of $^{60}$Ca+$^{60}$Ca  to 1.5 for
$^{60}$Ca+$^{40}$Ca, so the E$_{bal}$ also decreases. Now the
E$_{bal}$ for
$^{60}$Ca+$^{40}$Ca has same value as in case of symmetric reactions with
N/Z (N/A) = 1.5 (0.6) i.e. the value of E$_{bal}$ is decided by the N/Z (N/A) of the system and is independent of the asymmetry
of the reaction in agreement with \cite{supriya}.

\par
It has also been reported in literature that the MDI affects
drastically the collective flow as well as its disappearance \cite{soodmdi}. To
check the influence of MDI on the N/Z (N/A) dependence of E$_{bal}$ we
calculate the E$_{bal}$ for the whole N/Z (N/A) range from 1 to 2 (0.5 to 0.67) for the symmetric reactions with
SMD equation of state and symmetry potential F$_{1}(u)$. The
results are shown in figs. 5(c) and (d) (lower panels) by solid left triangles. We find that although the MDI
changes drastically the absolute value of E$_{bal}$ ( by about
30\%), however the N/Z (N/A) dependence of  E$_{bal}$ remains unchanged
on inclusion of MDI. Therefore, the dependence of  E$_{bal}$ as a
function of N/Z (N/A) on the symmetry energies of other different forms
(F$_{2}(u)$ and F$_{3}(u)$) is also expected to be preserved on
inclusion of MDI.

\section{Summary}
We have shown that the N/Z (N/A) dependence of E$_{bal}$ for
the isotopic series of Ca+Ca is a sensitive probe to
the symmetry energy as well as its density dependence at densities
higher than saturation density and is insensitive to other isospin
effects like Coulomb repulsion, and isospin dependence of nucleon-nucleon cross
section. We have also studied the effect of MDI on the N/Z (N/A) dependence of E$_{bal}$. We find
that although MDI influences the E$_{bal}$ drastically, the N/Z (N/A) dependence of E$_{bal}$ remains unchanged on inclusion of MDI.
\par
This work has been supported by a grant from Indo-French Centre
For The Promotion Of Advanced Research (IFCPAR) under project no.
4104-1.

\end{document}